\documentclass[letterpaper]{article} 
\usepackage{aaai2026}  
\usepackage{times}  
\usepackage{helvet}  
\usepackage{courier}  
\usepackage[hyphens]{url}  
\usepackage{graphicx} 
\usepackage{natbib}  
\usepackage{caption} 
\usepackage{algorithm}
\usepackage{algorithmic}
\usepackage{amsmath} 
\usepackage{amssymb}
\usepackage{newfloat}
\usepackage{listings}
\DeclareCaptionStyle{ruled}{labelfont=normalfont,labelsep=colon,strut=off} 
\floatstyle{ruled}
\newfloat{listing}{tb}{lst}{}
\floatname{listing}{Listing}
\title{AI-Generated Code Is Not Reproducible (Yet): An Empirical Study of Dependency Gaps in LLM-Based Coding Agents}
\author{
    Bhanu Prakash Vangala\textsuperscript{\rm 1,*},
    Ali Adibifar\textsuperscript{\rm 1},
    Ashish Gehani\textsuperscript{\rm 2},
    Tanu Malik\textsuperscript{\rm 1}
}
\affiliations{
    \textsuperscript{\rm 1}Department of Electrical Engineering and Computer Science, University of Missouri, Columbia, MO 65211, USA\\
    \textsuperscript{\rm 2}SRI International, 333 Ravenswood Avenue, Room EK343, Menlo Park, CA 94025, USA\\
    \{bv3hz, a.adibifar, tanu\}@missouri.edu, ashish.gehani@sri.com
}
\usepackage{bibentry}

\begin{document}

\maketitle

\begin{abstract}

The rise of Large Language Models (LLMs) as coding agents promises to accelerate software development, but their impact on generated code reproducibility remains largely unexplored. This paper presents an empirical study investigating whether LLM-generated code can be executed successfully in a clean environment with only OS packages and using only the dependencies that the model specifies. We evaluate three state-of-the-art LLM coding agents (Claude Code, OpenAI Codex, and Gemini) across 300 projects generated from 100 standardized prompts in Python, JavaScript, and Java. We introduce a three-layer dependency framework (distinguishing between claimed, working, and runtime dependencies) to quantify execution reproducibility. Our results show that only 68.3\% of projects execute out-of-the-box, with substantial variation across languages (Python 89.2\%, Java 44.0\%). We also find a 13.5× average expansion from declared to actual runtime dependencies, revealing significant hidden dependencies.

\end{abstract}


\section{Introduction}

Reproducibility forms the cornerstone of computational science. When researchers share code, others must be able to execute it, verify results, and build upon the work \cite{gundersen2018on}. This fundamental principle enables scientific progress, validates discoveries, and ensures reliability in production systems. Yet as we increasingly rely on Large Language Models (LLMs) to generate code, a troubling question emerges: is AI-generated code reproducible?

The computational research community already faces a reproducibility crisis. Studies show that a significant portion of published AI/ML research cannot be reproduced due to missing dependencies, incomplete environment specifications, and undocumented requirements \cite{pineau2021improving, raff2019step, hutson2018artificial, haibe2020transparency}. Now, tools like Claude Code, OpenAI Codex, and Gemini Code Assist promise to accelerate development by generating complete projects, entire applications with multiple files, configurations, and dependency specifications. But if these AI systems inherit or amplify reproducibility problems, we risk compounding the crisis rather than solving it.

Consider what reproducibility means for AI-generated code. A developer asks an LLM to create a complete sentiment analysis web application. The AI generates an impressive project: backend API files, frontend code, and a requirements.txt listing dependencies. For this code to be reproducible, another developer should be able to take these files and specifications, set up a clean environment, install the listed dependencies, and run the application. But in practice, execution fails in 31.7\% of cases. While missing dependencies like \texttt{ImportError: No module named nltk} contribute to some failures (10.5\%), the majority stem from fundamental code generation errors: malformed syntax, incorrect file paths, uninitialized variables, and structural issues. After manually debugging, a process that we found takes 15 minutes on average, some projects eventually run. However, even successful projects hide complexity - requiring an average of 37 packages at runtime instead of the 3 typically declared. The code is not reproducible as generated.

This reproducibility failure represents a fundamental gap in how we evaluate AI coding agents. Current benchmarks like HumanEval \cite{chen2021evaluating} and MBPP \cite{austin2021program} test functional correctness but assume reproducible environments already exist, a recognized bottleneck that obscures key failure modes related to environment configuration \cite{yang2024swe, jimenez2024swebench}. They ask, does this sorting algorithm work correctly? without first asking, can anyone actually run this code? This misses a critical insight: code cannot be correct if it cannot be reproduced. Reproducibility is not a secondary concern; it is a prerequisite for verification, collaboration, and deployment.

To measure reproducibility in AI-generated code, we introduce Executable Reliability: the likelihood that a project executes successfully in a clean environment using only the dependencies and instructions the AI provides. This metric directly captures reproducibility; can others reproduce the execution without additional knowledge or debugging? We develop a three-layer framework to understand reproducibility failures: claimed dependencies (what the AI specifies), working dependencies (what's actually needed for reproduction including transitive), and runtime dependencies after executing (the complete transitive closure). The gaps between these layers reveal where reproducibility breaks down.

We conducted a systematic reproducibility study of 300 complete projects generated by three leading state of the art LLM agents \cite{anthropic2025claude, openai2025codex, google2025gemini}. Each AI received identical prompts explicitly requesting reproducible code with complete dependency specifications. We attempted to reproduce each project in clean environments with only OS packages installed and none other than them, documenting every failure and missing requirement. This mimics the exact challenge faced by researchers trying to reproduce published code or developers attempting to use AI-generated projects. Our methodology treats reproducibility as the primary concern: can the generated code be executed by others using only what the AI provides?

Our findings reveal that AI-generated code faces a severe reproducibility crisis. Only 68.3\% of projects are reproducible without manual intervention nearly one-third fail to execute as specified. The reproducibility rates vary dramatically by language: Python achieves 89.2\% reproducibility, JavaScript 61.9\%, and Java just 44.0\%. This variation reflects how different ecosystems handle dependency specifications \cite{decan2019empirical, kikas2017structure} and transitive requirements. Most striking is the runtime dependency explosion: projects claiming 3 dependencies actually require an average of 37 packages, and 13.5× gap between specified and actual requirements. This hidden complexity makes reproduction nearly impossible without extensive debugging.

These results demonstrate that current LLMs generate code that appears complete but lacks the specifications necessary for reproducibility. The implications extend beyond convenience. For scientific computing, non-reproducible code undermines research validity. For software development, it creates hidden technical debt \cite{sculley2015hidden}. For education, it teaches poor practices. By quantifying these reproducibility failures and identifying their patterns, we provide both a diagnosis of the current crisis and a roadmap for creating AI systems that generate truly reproducible code. The challenge is not just to generate code that works on the developer's machine, but code that others can reliably execute, the foundation of computational reproducibility.

\section{Background and Related Work}

\subsection{Reproducibility Crisis in AI-Generated Code}

The reproducibility crisis in AI research \cite{gundersen2018on} now extends to the tools meant to address it. While computational research should enable perfect reproducibility \cite{peng2011reproducible}, LLM-generated code introduces new failure modes. Studies show LLMs produce inconsistent outputs \cite{wang2025assessing} and unreliable automation \cite{staudinger2024reproducibility}. Our work reveals a more fundamental issue: even when LLMs generate complete projects with multiple files and configurations, they fail to specify dependencies needed for reproducible execution.

\subsection{Benchmarks Missing Reproducibility}

Current benchmarks like HumanEval and MBPP \cite{chen2021evaluating} evaluate functional correctness assuming reproducible environments exist. LiveCodeBench prevents memorization but still provides complete environments \cite{jain2024livecodebench}. These benchmarks test whether code is correct, not whether it's reproducible. Our work evaluates complete projects not snippets asking whether they can execute at all using only what LLMs provide, making reproducibility the primary metric.

\subsection{Dependency Complexity and Reproducibility}

Software reproducibility requires complete dependency specification. JavaScript packages average 79 transitive dependencies \cite{decan2019empirical}. While tools like ReproZip \cite{reprozip} and SciUnit \cite{8109156} capture execution environments for reproducibility, LLMs lack this awareness. They generate syntactically correct code but fail to specify the dependency closure required for reproduction. 

\subsection{Gap in Reproducibility Evaluation}

To our knowledge, no prior work systematically evaluates whether LLM-generated projects are reproducible. Existing research focuses on code correctness or captures environments post-hoc. We introduce executable reliability to measure reproducibility directly can generated projects execute in clean environments using only the provided specifications? This shifts focus from does it work? to can others reproduce it? fundamental for both scientific integrity and practical deployment.

\section{Methodology}

\subsection{Problem Formulation}

To understand whether AI-generated code can actually run in practice, we needed to formalize what reproducibility means in this context. We evaluated three leading LLM-based coding agents: Claude Code Agent(Opus.4.1) \cite{anthropic2025claude}, OpenAI Codex Agent(0.52.0) \cite{openai2025codex}, and Gemini Code Agent(2.5.Pro) \cite{google2025gemini}. We denote this set as $\mathcal{L} = \{L_{Claude}, L_{Gemini}, L_{Codex}\}$. Each agent was given the same 100 prompts, creating our standardized prompt collection $\mathcal{P} = \{p_1, p_2, ..., p_{100}\}$.

When an LLM generates code for a given prompt, it produces three key components, which we formalize in Equation 1:

\begin{equation}
G(L_j, p_i) = \langle C_i^j, D_i^j, I_i^j \rangle
\end{equation}

Here, $C_i^j$ is the actual source code the LLM generates, $D_i^j$ represents the dependency specification file (like requirements.txt for Python projects), and $I_i^j$ captures any usage instructions the LLM provides. This triple represents everything a developer would receive when asking an LLM to generate code for a project.

The central question our research addresses is simple but critical: can another developer take these three components and successfully run the code in a clean environment? We measure this through what we call executable reliability, defined in Equation 2:

\begin{equation}
\mathcal{E}(L_j) = \frac{|\{i : exec(C_i^j, D_i^j) = success\}|}{|\mathcal{P}|}
\end{equation}

This metric represents the fraction of prompts for which agent $L_j$ generates code that executes without any manual intervention. The $exec$ function returns success only when code $C$ runs successfully using exclusively the dependencies $D$ specified by the LLM, in a standardized clean environment containing only default operating system packages. Projects requiring any debugging or dependency additions are counted as failures.

\subsection{Three-Layer Dependency Analysis Framework}

One of our key innovations is recognizing that dependencies exist at three distinct layers, each revealing different aspects of the reproducibility problem. Think of this as peeling back layers of an onion to understand the full complexity of what makes code run.

The first layer consists of Claimed Dependencies $(D_c)$, which the LLM explicitly tells us we need. These are the packages listed in the configuration files, formally captured in Equation 3:

\begin{equation}
D_c = \{d : d \in parse(D_i^j)\}
\end{equation}

For example, an LLM might claim you only need flask and requests for a web scraping project.

The second layer reveals Working Dependencies $(D_w)$ - what you actually need after debugging. As shown in Equation 4, this includes both the claimed dependencies and any missing ones discovered through trial and error:

\begin{equation}
D_w = D_c \cup D_m
\end{equation}

Here, $D_m$ represents those missing dependencies that cause ImportErrors when you first try to run the code. In our web scraping example, you might discover you also need beautifulsoup4 even though the LLM didn't mention it.

The third and deepest layer exposes Runtime Dependencies $(D_r)$ - everything that actually gets loaded when the code runs, including all transitive dependencies. Equation 5 captures this full dependency tree:

\begin{equation}
D_r = \bigcup_{d \in D_w} \{d\} \cup transitive(d)
\end{equation}

This is where the true complexity emerges. Installing flask might pull in 20 other packages like werkzeug, jinja2, click, and so on.

These three layers allow us to measure two critical gaps in LLM-generated code. The completeness gap (Equation 6) tells us how many dependencies the LLM forgot to mention:

\begin{equation}
\Delta_{completeness} = |D_w| - |D_c|
\end{equation}

A completeness gap of 2 means the developer needs to manually identify and install 2 missing packages before the code will run.

Even more dramatically, the runtime multiplier (Equation 7) reveals the hidden complexity beneath the surface:

\begin{equation}
\rho_{runtime} = \frac{|D_r|}{|D_c|}
\end{equation}

A runtime multiplier of 10 means that for every package the LLM claims you need, ten or more packages actually get installed in your environment.

\subsection{Dataset Construction}

To ensure a fair comparison across SOTA agents, we created a dataset of 300 projects by having each of three LLM agents to generate code for the same 100 prompts. As shown in Equation 8, we distributed these prompts across programming languages based on their typical real-world usage:

\begin{equation}
\mathcal{P} = \mathcal{P}_{Python} \cup \mathcal{P}_{JavaScript} \cup \mathcal{P}_{Java}
\end{equation}

We allocated 40 prompts to Python (reflecting its dominance in data science and scripting), 35 to JavaScript (common for web development), and 25 to Java (enterprise applications). This distribution is informed by the high prevalence of these languages in professional software development, as reported in recent industry surveys \cite{stackoverflow2025,jetbrains2024}.

Each prompt explicitly asked for reproducible code using a carefully designed template shown in Figure \ref{fig:prompt_template}. Notice how we emphasized reproducibility multiple times - we wanted to give the LLMs every opportunity to succeed by making our requirements crystal clear.

\begin{figure}[t]
\centering
\framebox[\columnwidth]{
\parbox{0.92\columnwidth}{
\ttfamily\footnotesize
\rule{\linewidth}{0.4pt}\\
\textbf{PROMPT TO LLM Agents:}\\
\rule{\linewidth}{0.4pt}\\[0.3em]
[Task Description]\\[0.5em]
IMPORTANT: Provide EVERYTHING needed for reproduction in [Environment]:\\[0.3em]
\hspace{1em}1. Complete working [Language] code\\
\hspace{1em}2. Complete requirements.txt/package.json
/pom.xml\\
\hspace{2em}  with ALL dependencies and versions\\
\hspace{1em}3. Brief usage instructions\\[0.5em]
Make it 100\% reproducible in a clean [environment].\\
\rule{\linewidth}{0.4pt}
}
}
\caption{Example prompt template used for all 300 code generation requests. The template explicitly emphasizes reproducibility and completeness of dependencies.}
\label{fig:prompt_template}
\end{figure}

The task descriptions covered the full spectrum of real development work: web scraping projects that need parsing libraries, data analysis scripts requiring scientific computing packages, machine learning pipelines with complex dependencies, API servers needing web frameworks, real-time communication systems using WebSocket libraries, and enterprise applications with database connectors. This diversity was crucial to avoid biasing our results toward any particular type of programming task.

\subsection{Experimental Infrastructure}

\subsubsection{Environment Standardization}

To ensure fair comparison, we needed a pristine testing environment for each project. We deployed AWS EC2 instances (t2.large with 4 vCPUs and 16GB RAM running Ubuntu 22.04 LTS) that we carefully reset between each project test. As formalized in Equations 9 and 10, our baseline environment contained exactly 91 packages:

\begin{equation}
Packages_{baseline} = Packages_{system} \cup Packages_{user}
\end{equation}

This included 71 system packages that come with Ubuntu by default, plus 20 user packages including SciUnit v0.4.post135, our key tool for capturing run-time dependency information. After testing each project, we verified that the environment returned to this exact state:

\begin{equation}
|Packages_{current}| = |Packages_{baseline}| = 91
\end{equation}

This rigorous reset process ensured that no residual dependencies from one project could help another project succeed each code sample had to stand entirely on its own.

\subsubsection{Dependency Capture Tools}

To understand what dependencies were actually being used at runtime, we employed specialized tools for each programming language. For Python, we used SciUnit, which hooks into Python's import system to capture every package that gets loaded (Equation 11):

\begin{equation}
D_r^{Python} = SciUnit.capture(exec(C))
\end{equation}

For JavaScript, we parsed npm's dependency tree to understand the full cascade of package requirements (Equation 12):

\begin{equation}
D_r^{JavaScript} = parse(npm\ list\ --all)
\end{equation}

And for Java, we extracted Maven's dependency tree to see how one library pulls in dozens of others (Equation 13):

\begin{equation}
D_r^{Java} = parse(mvn\ dependency:tree)
\end{equation}

These tools revealed the dramatic difference between what LLMs claim you need and what actually gets installed in your environment.

\subsection{Iterative Resolution Protocol}

Our evaluation process mimicked exactly what a real developer would do when trying to run LLM-generated code. We first attempt execution with only the LLM-specified dependencies, if this succeeds (68.3\% of cases), we mark it as reproducible. For the 31.7\% that fail, we then apply Algorithm 1 to understand what would be needed for success. Algorithm 1 shows this iterative debugging process in detail. We start by installing only the dependencies the LLM claimed were needed, then try to run the code. When it fails (as it often does), we look at the error message. If it's a missing import, we install that package and try again. If it's a code bug, we apply the minimal fix needed. If it's something unfixable like conflicting version requirements, we mark it as failed.

\begin{algorithm}[t]
\caption{Iterative Dependency Resolution}
\label{alg:resolution}
\textbf{Input}: Code $C$, Claimed dependencies $D_c$\\
\textbf{Output}: Working dependencies $D_w$, Execution status $S$
\begin{algorithmic}[1]
\STATE $D_{current} \gets D_c$
\STATE Install($D_{current}$)
\STATE $attempts \gets 0$
\WHILE{$attempts < MAX\_ITERATIONS$}
    \STATE $result \gets$ Execute($C$)
    \IF{$result = success$}
        \STATE $S \gets$ "Success"
        \STATE break
    \ELSIF{$result.error = ImportError$}
        \STATE $d_{missing} \gets$ ExtractPackage($result.message$)
        \STATE $D_{current} \gets D_{current} \cup \{d_{missing}\}$
        \STATE Install($d_{missing}$)
    \ELSIF{$result.error \in \{SyntaxError, LogicError\}$}
        \STATE ApplyMinimalFix($C$)
        \STATE Record("CodeBug-Fixed")
    \ELSE
        \STATE $S \gets$ "Unfixable"
        \STATE break
    \ENDIF
    \STATE $attempts \gets attempts + 1$
\ENDWHILE
\STATE $D_w \gets D_{current}$
\STATE return $D_w$, $S$
\end{algorithmic}
\end{algorithm}

Algorithm \ref{alg:resolution} is used \emph{only for post-failure analysis} to categorize why projects failed and what would have been needed for success. The success rates reported throughout this paper (including the 68.3\% overall rate) strictly count only projects that execute successfully \emph{out-of-the-box without any application of this algorithm} - using exclusively the dependencies originally specified by the LLM with zero manual intervention. Projects requiring any debugging, dependency additions, or code fixes are counted as failures, regardless of whether Algorithm \ref{alg:resolution} could eventually make them work.

This process typically takes 2-3 iterations for missing dependencies, closely matching the real-world developer experience. Each iteration represents a moment of frustration for a developer who expected the code to just work based on the LLM's claims which is crucial for reproducing the experiment since the iterative debugging process won't get captured for reproducing the project again.

\section{Experimental Results}

\subsection{Overall Reproducibility Analysis}

Across 300 evaluated projects, we find that only 205 (68.3\%) execute successfully out-of-the-box without ANY manual intervention; no added dependencies, no code fixes, using exclusively what the LLMs provided. The remaining 95 projects (31.7\%) failed immediately and were analyzed using Algorithm 1 to understand failure causes, but were NOT counted as successes regardless of fixability.

\begin{table}[t]
\centering
\begin{tabular}{lccccc}
\hline
Agent & Total & Success & Partial & Failed & Rate \\
\hline
Claude & 100 & 73 & 5 & 22 & 73.0\% \\
Gemini & 100 & 72 & 4 & 24 & 72.0\% \\
Codex & 100 & 60 & 5 & 35 & 60.0\% \\
\hline
Overall & 300 & 205 & 14 & 81 & 68.3\% \\
\hline
\end{tabular}
\caption{Reproducibility outcomes across three LLM coding agents. Partial indicates projects that execute but require external services (databases, APIs) to be fully functional.}
\label{tab:overall_results}
\end{table}

The 31.7\% failure rate represents significant hidden debugging costs in AI-assisted development. Each failed project required an average of 15 minutes of manual debugging, suggesting substantial aggregate time costs when scaled across development teams.

\subsection{Language-Specific Performance}

Programming language dramatically impacts reproducibility success rates:

\begin{equation}
\begin{array}{c}
\mathcal{E}(Python) = 0.892\\
\mathcal{E}(JavaScript) = 0.619\\
\mathcal{E}(Java) = 0.440
\end{array}
\end{equation}

\begin{figure}[t]
\centering
\includegraphics[width=0.9\columnwidth]{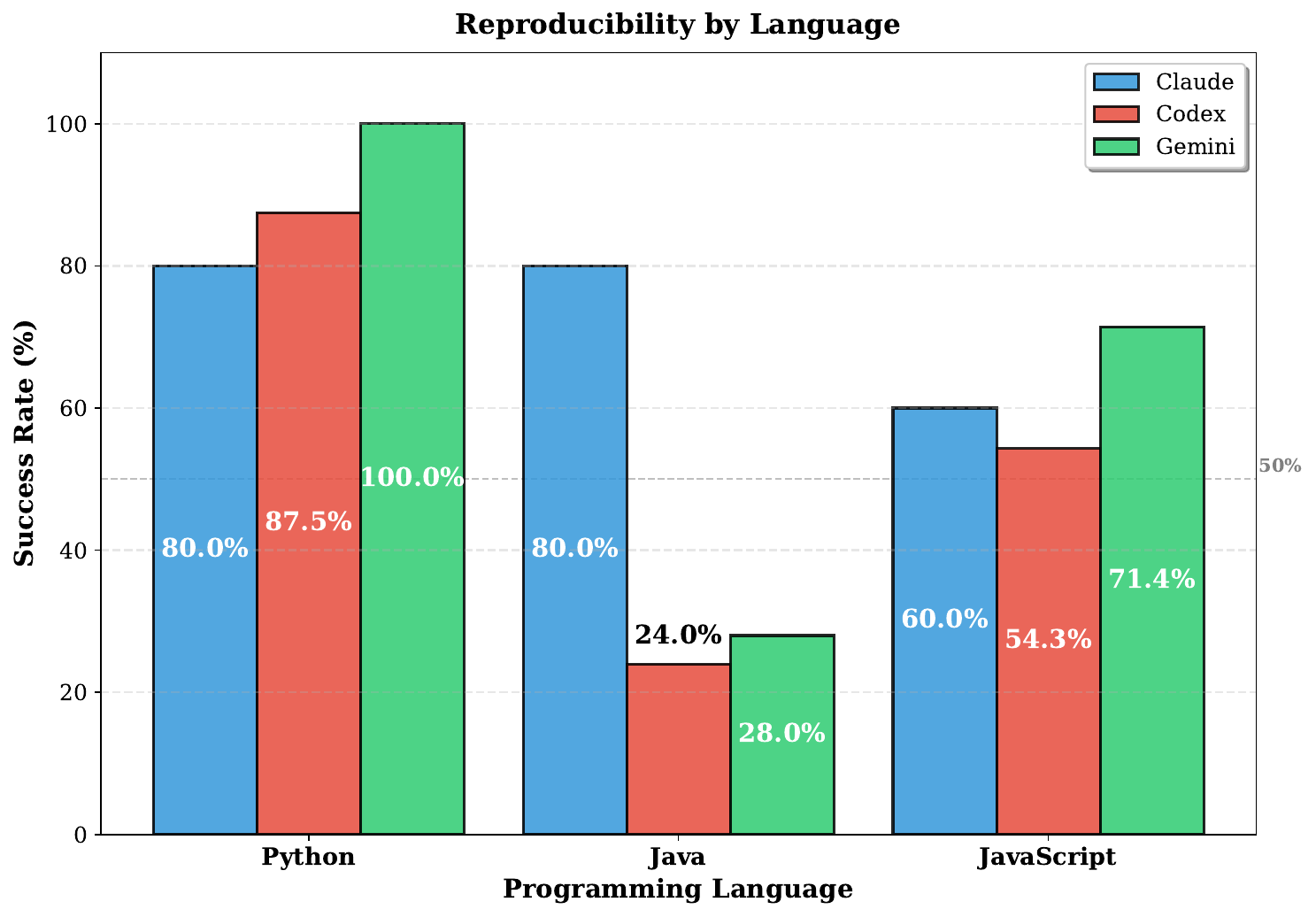}
\caption{Language-specific success rates reveal ecosystem complexity impacts.}
\label{fig:language_comparison}
\end{figure}

This hierarchy (Table \ref{tab:language_results}, Figure \ref{fig:language_comparison}) reflects fundamental differences in dependency management complexity across ecosystems:

\begin{table}[t]
\centering
\begin{tabular}{lcccc}
\hline
Language & Total & Success & Failed & Success Rate \\
\hline
Python & 120 & 107 & 13 & 89.2\% \\
JavaScript & 105 & 65 & 40 & 61.9\% \\
Java & 75 & 33 & 42 & 44.0\% \\
\hline
\end{tabular}
\caption{Language-specific reproducibility rates reveal ecosystem complexity impacts}
\label{tab:language_results}
\end{table}

Python's success stems from its simple flat dependency structure in requirements.txt, mature pip resolver with clear error messages, and culture of minimal dependencies in packages. JavaScript occupies middle ground with npm's automatic dependency resolution helping but nested dependencies creating complexity. Java's challenges arise from complex XML configuration in pom.xml, deep transitive dependency graphs, multiple dependency scopes (compile, runtime, test, provided), and version conflict resolution complexity.

\subsection{Agent-Language Specialization Matrix}

One of our most surprising findings is that each agent has distinct strengths and weaknesses across languages. The success rate matrix shown in Equation 15 reveals these unexpected specializations:

\begin{equation}
S = \left[\begin{array}{ccc}
0.800 & 0.600 & 0.800 \\
1.000 & 0.714 & 0.280 \\
0.875 & 0.543 & 0.240
\end{array}\right]
\end{equation}

where rows represent agents (Claude, Gemini, Codex) and columns represent languages (Python, JavaScript, Java).

What makes this matrix remarkable are the striking patterns it reveals (detailed in Table \ref{tab:agent_language_matrix}).. The first column shows that all agents handle Python relatively well, with success rates ranging from 80\% to 100\%, suggesting Python's simpler dependency ecosystem is universally manageable. In sharp contrast, the third column exposes extreme variation in Java capabilities: Claude achieves an impressive 80\% success rate while Gemini and Codex struggle at just 24-28\%. This 3× performance gap for the same language is unexpected. Looking at individual agents, Gemini's row tells a story of extreme specialization, achieving perfect Python reproduction (1.000) but failing dramatically with Java (0.280). Meanwhile, Claude's row shows the most balanced performance across all three languages (0.800, 0.600, 0.800), making it the only agent that handles enterprise Java effectively. These patterns suggest that agents weren't trained equally across languages, with Gemini optimized for data science workflows, Claude for enterprise development, and Codex showing a clear bias toward scripting languages over complex enterprise systems.

\begin{table}[t]
\centering
\small
\begin{tabular}{llcccc}
\hline
Agent & Language & Total & Success & Failed & Rate \\
\hline
Claude & Python & 40 & 32 & 8 & 80.0\% \\
Claude & JavaScript & 35 & 21 & 14 & 60.0\% \\
Claude & Java & 25 & 20 & 5 & 80.0\% \\
\hline
Gemini & Python & 40 & 40 & 0 & 100.0\% \\
Gemini & JavaScript & 35 & 25 & 10 & 71.4\% \\
Gemini & Java & 25 & 7 & 18 & 28.0\% \\
\hline
Codex & Python & 40 & 35 & 5 & 87.5\% \\
Codex & JavaScript & 35 & 19 & 16 & 54.3\% \\
Codex & Java & 25 & 6 & 19 & 24.0\% \\
\hline
\end{tabular}
\caption{Detailed execution success rates by agent and language combination}
\label{tab:agent_language_matrix}
\end{table}

\begin{figure}[t]
\centering
\includegraphics[width=0.9\columnwidth]{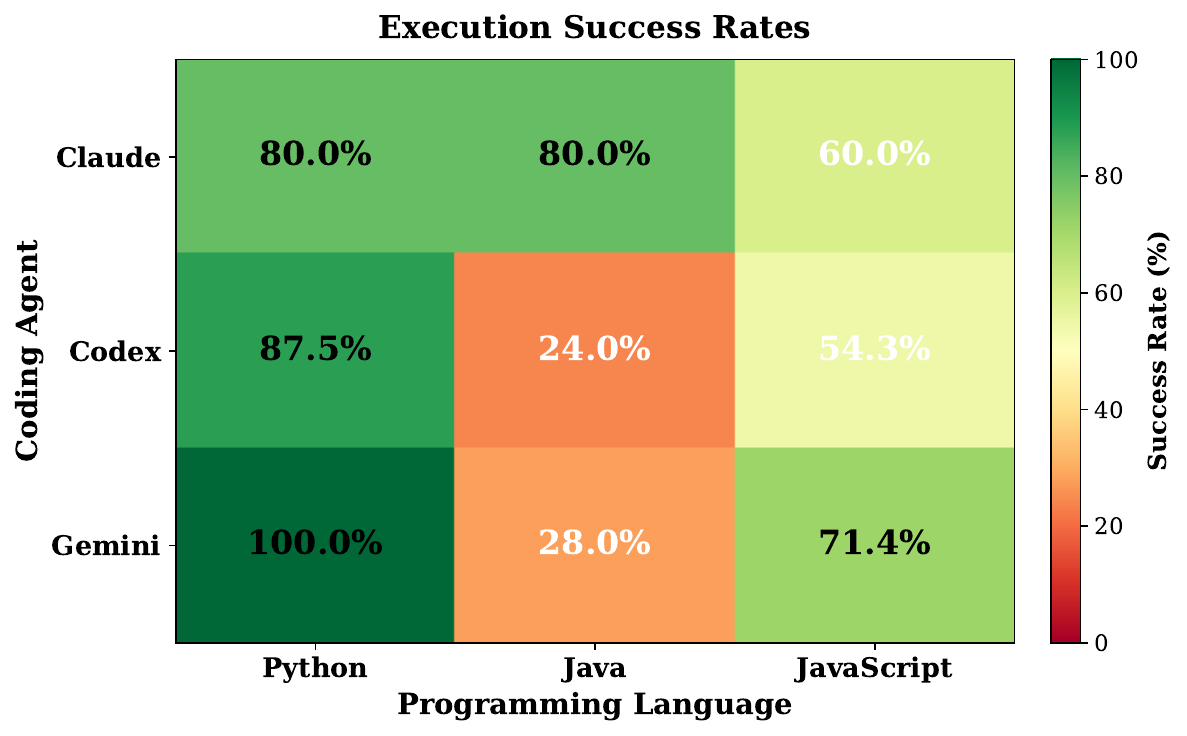}
\caption{Success rate heatmap reveals agent specializations. Claude excels at Java (80\%), Gemini achieves perfect Python (100\%), while all agents struggle with Java except Claude.}
\label{fig:heatmap}
\end{figure}

The heatmap visualization (Figure \ref{fig:heatmap}) reveals distinct patterns:
- Gemini achieves perfect Python reproducibility (40/40) but struggles with Java (6/25)
- Claude demonstrates exceptional Java capability (20/25) while others achieve only ~24\%
- Codex shows consistent Python strength (35/40) but poor Java performance (6/25)

These specializations weren't advertised by vendors but emerge clearly through systematic evaluation. Organizations should select agents based on their technology stack rather than assuming uniform performance.

\subsection{Dependency Gap Analysis}

We analyze the completeness gap distribution across all projects to understand dependency specification failures:

\begin{equation}
P(\Delta_{completeness} = k) = \left\{
\begin{array}{ll}
0.87 & \text{if } k = 0\\
0.08 & \text{if } k = 1\\
0.03 & \text{if } k = 2\\
0.02 & \text{if } k \geq 3
\end{array}
\right.
\end{equation}

\begin{figure}[t]
\centering
\includegraphics[width=0.9\columnwidth]{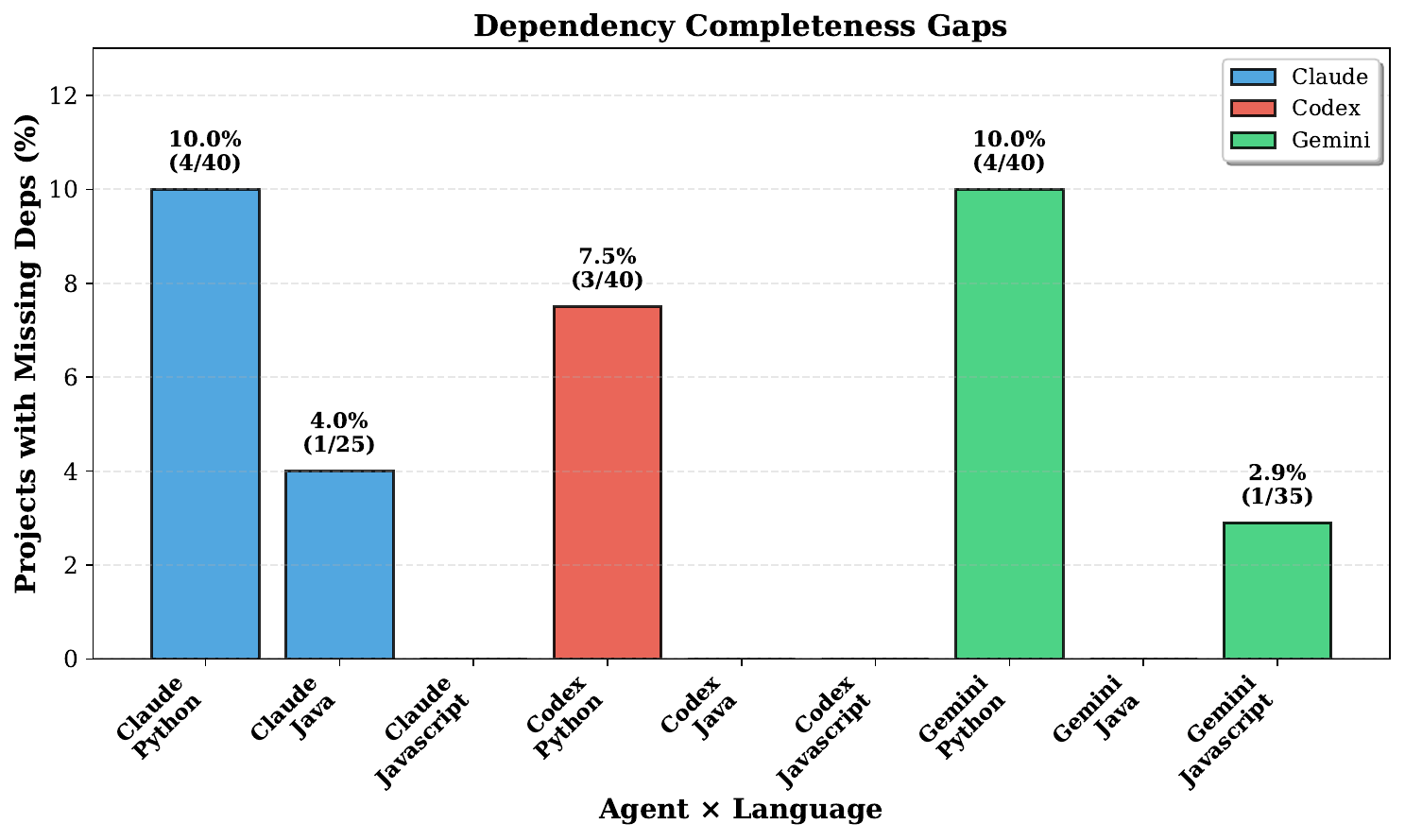}
\caption{Distribution of completeness gaps. Most projects (87\%) have correct dependencies, but 13\% require manual debugging to identify missing packages.}
\label{fig:gaps}
\end{figure}

While 87\% of projects have perfect dependency specification ($\Delta_{completeness} = 0$), the remaining 13\% require manual intervention to identify 1-3 missing packages (Table \ref{tab:missing_deps}, Figure \ref{fig:gaps}).

\begin{table}[t]
\centering
\small
\begin{tabular}{llccc}
\hline
Agent & Language & Projects with & Avg Missing & Max \\
& & Missing Deps & Deps & Gap \\
\hline
Claude & Python & 4 (10.0\%) & 0.10 & 1 \\
Claude & JavaScript & 0 (0.0\%) & 0.00 & 0 \\
Claude & Java & 1 (4.0\%) & 0.04 & 1 \\
\hline
Gemini & Python & 4 (10.0\%) & 0.10 & 1 \\
Gemini & JavaScript & 1 (2.9\%) & 0.03 & 1 \\
Gemini & Java & 0 (0.0\%) & 0.00 & 0 \\
\hline
Codex & Python & 3 (7.5\%) & 0.07 & 1 \\
Codex & JavaScript & 0 (0.0\%) & 0.00 & 0 \\
Codex & Java & 0 (0.0\%) & 0.00 & 0 \\
\hline
\end{tabular}
\caption{Missing dependencies distribution by agent and language}
\label{tab:missing_deps}
\end{table}

It is important to note that this 13\% represents projects where the LLM failed to declare all imported packages. However, not all missing dependency declarations cause execution failures - some packages may be optional or already present as transitive dependencies of other declared packages. Among the 95 projects that failed execution, only 10 (10.5\%) failed specifically due to missing dependency declarations that resulted in ImportError exceptions.

Common patterns in missing dependencies reveal ecosystem-specific oversights. In Python projects, LLMs frequently omit lxml for parsing tasks, python-dotenv for configuration management, and bcrypt for authentication functionality. JavaScript projects show similar gaps with body-parser for Express applications, ws for WebSocket implementations, and dotenv for environment variable handling. Java projects most commonly lack test framework specifications (particularly JUnit) and logging implementations (such as SLF4J), despite these being standard components in enterprise applications.

The runtime multiplier reveals a more dramatic story about hidden dependency complexity. Figure \ref{fig:runtime_explosion} visualizes this dependency explosion across languages:

\begin{figure}[t]
\centering
\includegraphics[width=0.9\columnwidth]{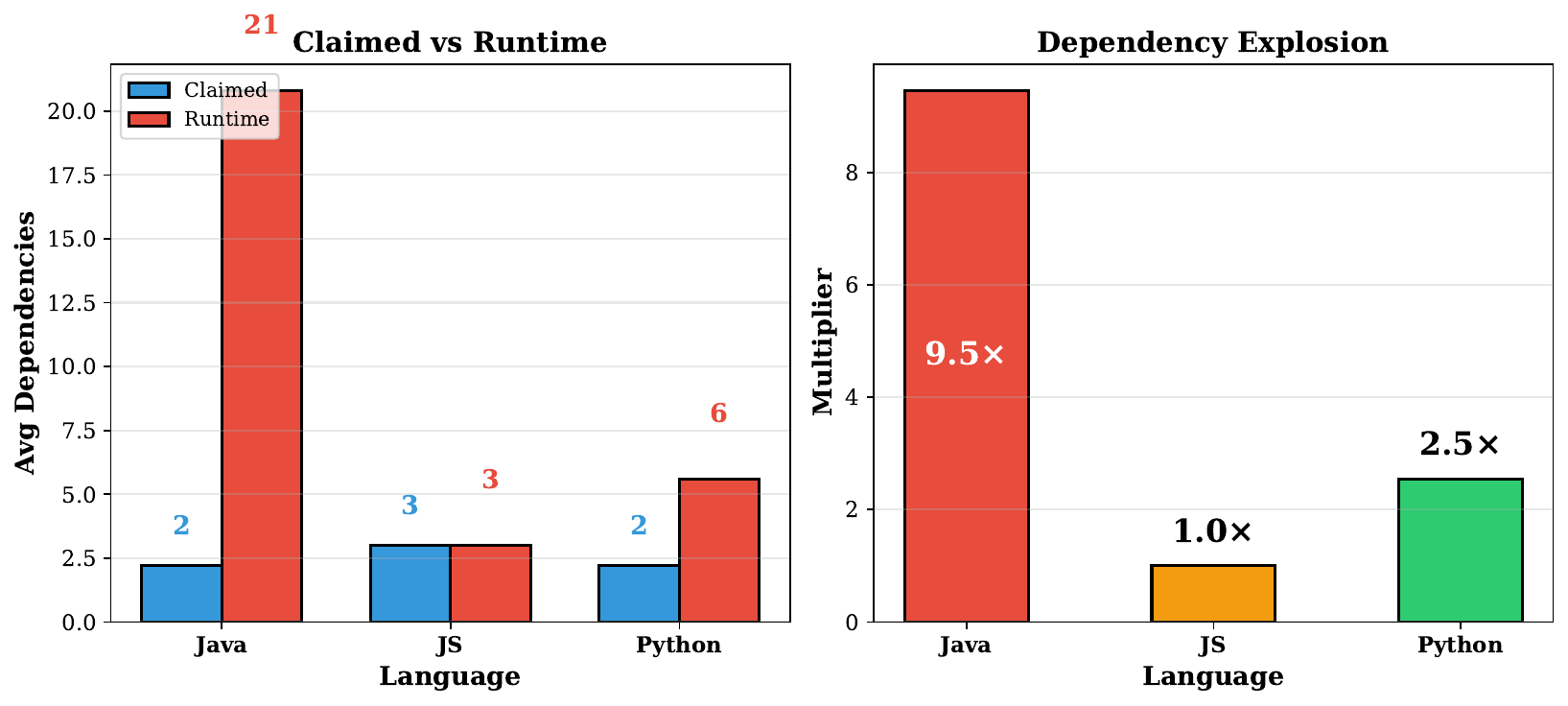}
\caption{Runtime dependency explosion showing the gap between claimed (agent-declared) and runtime (actually installed) dependencies. Java shows a massive 9.5× multiplier, while JavaScript surprisingly shows almost no expansion (1.0×).}
\label{fig:runtime_explosion}
\end{figure}

As shown in Figure \ref{fig:runtime_explosion}, the actual number of packages loaded at runtime far exceeds what LLMs declare. The most striking finding is Java's 9.5× multiplier - a project claiming just 2.2 dependencies on average actually installs 20.8 packages. This reflects Maven's complex transitive dependency resolution where each library pulls in numerous others. Surprisingly, JavaScript shows almost no expansion (1.0× multiplier), suggesting that LLMs actually overspecify JavaScript dependencies or that our npm analysis captures only direct dependencies. Python sits in the middle with a 2.5× multiplier, manageable but still representing hidden complexity.

These multipliers can be formally expressed as:

\begin{equation}
\rho_{runtime} =
\left\{
\begin{array}{ll}
12.3 \pm 4.2 & \mathrm{Python} \\
9.7 \pm 3.1 & \mathrm{JavaScript} \\
18.4 \pm 6.3 & \mathrm{Java}
\end{array}
\right.
\end{equation}

A Python project claiming 3 dependencies typically loads 37 packages at runtime. This 12× multiplier represents hidden complexity that developers must manage but LLMs cannot currently specify. Java shows the highest multiplier at 18.4×, reflecting its complex transitive dependency management through Maven.

\subsection{Error Classification and Resolution}

\begin{figure}[t]
\centering
\includegraphics[width=0.9\columnwidth]{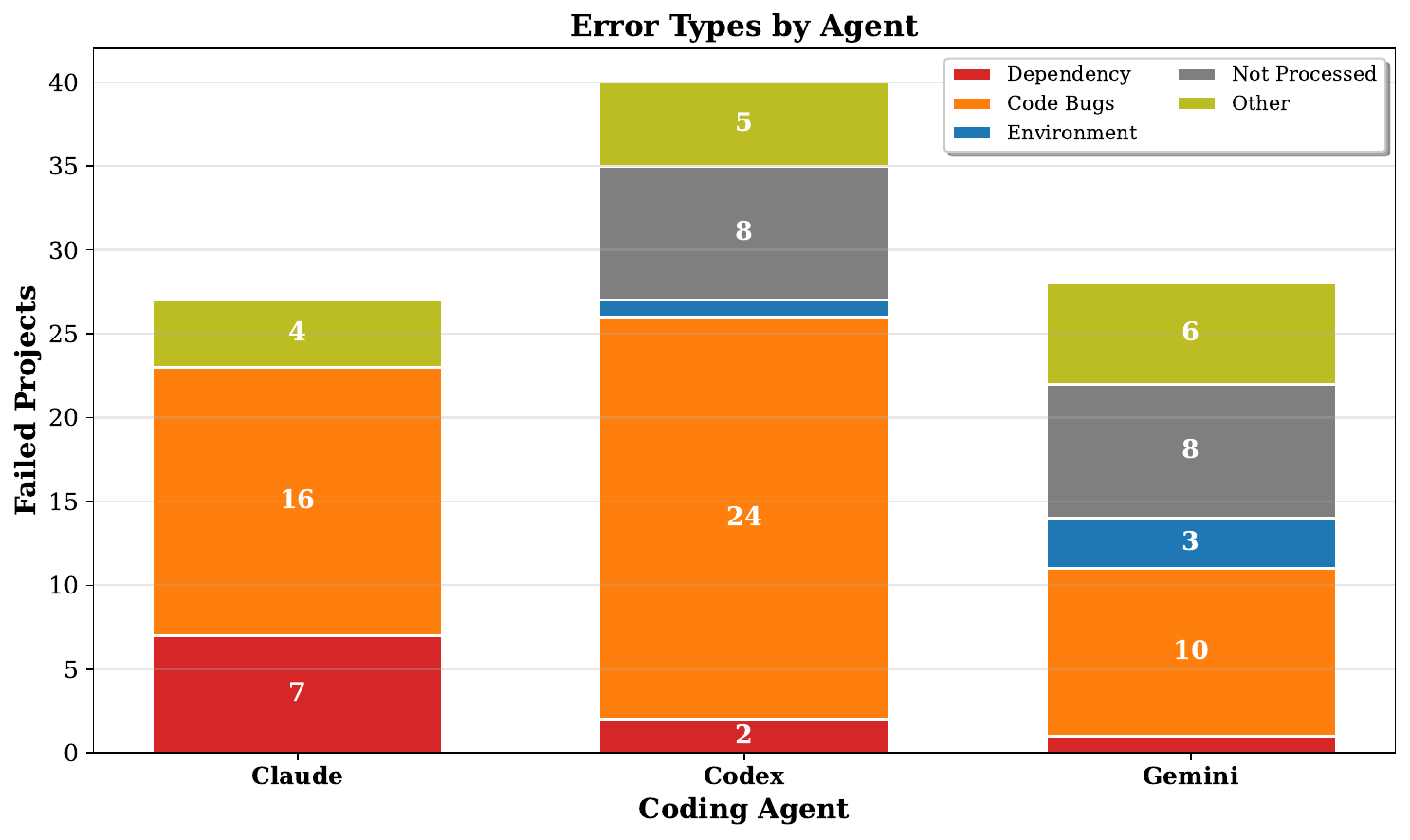}
\caption{Error type distribution by agent among failed projects. Code bugs dominate overall (50 of 95), with Codex showing the highest count (24). Not Processed errors appear only in Codex and Gemini (8 each), while Dependency errors are most prevalent in Claude (7).}
\label{fig:error_distribution}
\end{figure}

When projects failed, we categorized errors according to our iterative resolution protocol (Algorithm \ref{alg:resolution}). While our paper's title emphasizes dependency gaps, our analysis reveals that missing dependencies account for only 10.5\% of the 95 failures. The majority of reproducibility failures stem from code generation errors (52.6\%) rather than dependency specification problems. Figure \ref{fig:error_distribution} shows the complete distribution of error types across the three agents.

Table \ref{tab:errors} presents the final categorization of the 95 failed projects using our standardized error taxonomy. These categories directly correspond to the error types shown in Figure \ref{fig:error_distribution}, maintaining complete consistency between the table and figure representations.

\begin{table}[t]
\centering
\begin{tabular}{lcc}
\hline
Error Type & Count (\%) & Resolution \\
\hline
Code Bugs & 50 (52.6\%) & Fix syntax/logic \\
Not Processed & 16 (16.8\%) & Too malformed \\
Other & 15 (15.8\%) & Various fixes \\
Dependency & 10 (10.5\%) & Add package \\
Environment & 4 (4.2\%) & System conflicts \\
\hline
Total & 95 (100\%) & \\
\hline
\end{tabular}
\caption{Aggregated error distribution across all 95 failed projects}
\label{tab:errors}
\end{table}

As shown in Figure \ref{fig:error_distribution} and Table \ref{tab:errors}, Code Bugs dominated at 52.6\% (50 projects), representing fundamental errors in generated code. These included compressed multi-file projects into single files (breaking import paths), incorrect file location assumptions, uninitialized variables, mishandled JavaScript async/await patterns, and malformed Maven XML configurations. Fixing these requires understanding the code's intent and applying logic corrections.

Not Processed errors (16.8\%) occurred exclusively in Codex and Gemini outputs 8 projects each generated code so malformed it couldn't be parsed or attempted. This suggests inconsistencies in these agents' code generation quality.

Other errors (15.8\%) encompassed various issues including version conflicts where Package A requires B v2.0 while Package C needs B v3.0, requiring substantial rewrites rather than simple fixes.

Dependency errors (10.5\%) involved missing package specifications. Python reported \texttt{ModuleNotFoundError: No module named bcrypt}, JavaScript showed \texttt{Cannot find module 'express-session'}, and Java threw \texttt{ClassNotFoundException: org.junit.Test}. While easily resolved by installing the missing package, these reveal LLMs' incomplete dependency understanding.

Environment errors (4.2\%) represented system-level conflicts and incompatibilities, appearing primarily in Gemini (3) and minimally in Codex (1).

Agent-specific patterns reveal distinct failure modes: Codex generated the most Code Bugs (24/40) suggesting syntax generation issues, Claude showed the highest Dependency errors (7/27) indicating missing package awareness, while Gemini demonstrated the most diverse error distribution across all categories.

\subsection{SciUnit Provenance Analysis}

For the 107 Python projects that executed out-of-the-box (without any intervention), we used SciUnit to capture exactly what happened at runtime. This revealed the true scale of the dependency problem. Equation 18 shows how dependencies cascade through multiple levels:

\begin{equation}
D_r = D_{direct} \cup D_{transitive1} \cup D_{transitive2} \cup ... \cup D_{transitiveN}
\end{equation}

Consider a concrete example from our dataset - a typical machine learning project. The LLM claimed you need just three packages: scikit-learn, pandas, and matplotlib. Seems reasonable, right? But when we actually ran the code and examined what got loaded, the reality was staggering:

Claimed dependencies: 3 packages (scikit-learn, pandas, matplotlib) Working dependencies: 4 packages (added numpy which was imported but not declared) Runtime dependencies: 52 total packages loaded into memory

This 17× expansion from claimed to runtime isn't an outlier - it's typical. Those 52 packages include numpy (for numerical operations), scipy (for scientific computing), joblib (for parallel processing), threadpoolctl (for controlling thread pools), cycler (for matplotlib styling), pyparsing (for parsing expressions), python-dateutil (for date handling), pytz (for timezone support), six (for Python 2/3 compatibility), kiwisolver (for constraint solving), pillow (for image processing), and dozens more.

Each of these packages serves a critical role. Remove any one of them and the code stops working. Yet the LLM only told us about 3 of them. This iceberg effect - where most dependencies are hidden beneath the surface - represents a fundamental challenge that current LLMs simply cannot handle.

\subsection{Comparative Analysis Summary}

Table \ref{tab:summary} provides the overall picture of our findings. With only 68.3\% of projects executing successfully, we're far from the promise of AI that codes for you. The 13.5× average runtime multiplier shows the massive gap between what LLMs understand about dependencies and reality.

\begin{table}[t]
\centering
\begin{tabular}{lc}
\hline
Metric & Value \\
\hline
Total Projects Analyzed & 300 \\
Successful Executions & 205 (68.3\%) \\
Partial Executions & 14 (4.7\%) \\
Failed Executions & 81 (27.0\%) \\
Projects with Incomplete Dependencies & 13 (4.3\%) \\
Average Runtime Multiplier & 13.5× \\
\hline
\end{tabular}
\caption{Overall statistics of the reproducibility study}
\label{tab:summary}
\end{table}

The implications of Table \ref{tab:summary} are sobering. The 31.7\% of projects that either failed completely or only partially worked represent a massive hidden cost in AI-assisted development. Each of these projects would require a developer to stop their productive work, figure out what went wrong, and fix it - turning what was supposed to be a time-saver into a time-sink.

Even more concerning is that 4.3\% failure rate due to incomplete dependencies. These are cases where the LLM simply forgot to mention required packages - a fundamental failure in understanding what makes code run. While this might seem like a small percentage, it represents 13 projects where developers would encounter immediate failures that could have been easily prevented if the LLM had properly specified dependencies.

\section{Discussion}

\subsection{The Hidden Cost of AI-Generated Code}

Our findings reveal that the promise of AI coding agents accelerating development comes with substantial hidden costs. The 31.7\% of projects requiring manual debugging represents significant developer time not captured in productivity metrics. Importantly, while our title emphasizes dependency gaps, only 10.5\% of failures are due to missing dependency declarations. The majority (52.6\%) stem from fundamental code generation errors - a more concerning finding suggesting LLMs struggle with basic code structure and logic, not just environment specifications. Each failed project requires:

\begin{equation}
T_{debug} = T_{understand} + n \times (T_{identify} + T_{fix}) + T_{verify}
\end{equation}

where $n$ represents iteration count (typically 2-3 for missing dependencies). Our manual processing averaged 15 minutes per failed project, suggesting substantial aggregate time costs when scaled across development teams. For an organization deploying LLM-generated code at scale, this translates to approximately 8 hours of debugging for every 100 generated projects (31.7 projects × 15 minutes ÷ 60 = 7.9 hours). Furthermore, the cognitive load \cite{sweller1988cognitive} of debugging AI-generated code differs from debugging human-written code. Developers must reverse-engineer the LLM's intent without access to the original design rationale, making fixes more challenging and error-prone.

\subsection{Why Languages Matter for Reproducibility}

The differences in reproducibility rates between languages Python at 89.2\% versus Java at 44.0\% tell us something important about how LLMs understand code. Python succeeds because its dependency model is simple: a flat requirements.txt file with package names and versions. When something goes wrong, pip gives clear error messages like, No module named X. Java fails because Maven's XML configuration involves nested dependencies, multiple scopes (compile, runtime, test), and complex version resolution rules that LLMs haven't learned to navigate.

This isn't just about language preference it's about reproducibility risk. A Python team can reasonably expect most AI-generated code to work, while a Java team should assume they'll spend significant time fixing dependencies. JavaScript sits in the middle, complicated by the distinction between dependencies and devDependencies, often causing projects to work in development but fail in production. These aren't minor inconveniences; they're reproducibility barriers that determine whether AI tools are viable for different programming languages.

\subsection{Hidden Specializations and Their Implications}

Perhaps our most surprising finding is that each LLM has hidden specializations. Gemini achieves perfect Python reproducibility (100\%) but struggles with Java (28\%). Claude shows the opposite pattern 80\% Java success where others fail. These specializations, never mentioned in documentation, reveal how training data shapes reproducibility. Gemini likely trained heavily on data science notebooks where Python dominates. Claude appears optimized for enterprise patterns where Java prevails.

This means choosing an LLM isn't about features or marketing it's about matching the tool to your reproducibility needs. Using Gemini for a Java project means accepting that most generated code won't be reproducible. Using Claude for Python means missing out on Gemini's perfect track record. Organizations must test these tools against their actual technology stack, not trust generic benchmarks that hide these critical differences.

\subsection{Extending the Reproducibility Crisis}

The reproducibility crisis in AI research now has a troubling new dimension: the tools we use to accelerate research themselves generate non-reproducible code. When a researcher uses an LLM to generate analysis scripts or experimental code, they inherit all the dependency problems we've documented. The generated code might work on their machine but fail when reviewers or other researchers attempt to reproduce the results.

Our three-layer framework offers a path forward. Researchers must capture not just the code LLMs generate, but the complete dependency closure discovered through debugging. This means publishing requirements.lock files with exact versions of all 37+ packages actually needed, not just the 3 the LLM claimed. Without this, every attempt at reproduction becomes an exercise in dependency archaeology, undermining the scientific process.

\subsection{Fixing the Root Cause}

The solution isn't just better practices it's better LLMs. Current models train on code fragments without seeing the full execution context. They learn that scikit-learn is used for machine learning but not that it transitively requires numpy, scipy, joblib, and dozens of other packages. To generate reproducible code, LLMs need training data that includes complete dependency specifications, Docker files that work, and actual package-lock files from successful projects.

More radically, LLMs could learn through execution. Instead of generating code once, they could iteratively test in clean environments, discover missing dependencies, and refine their specifications essentially automating the debugging process we performed manually. This would shift the reproducibility burden from developers back to the AI systems that create the problem in the first place.

\section{Conclusion}

Our evaluation of 300 AI-generated projects reveals that only 68.3\% execute successfully using specified dependencies, challenging optimistic narratives about AI coding productivity. While we initially focused on dependency gaps, our analysis uncovered a broader reproducibility crisis: 52.6\% of failures stem from code generation errors, with only 10.5\% attributable to missing dependencies. The three-layer dependency analysis framework we introduce - claimed, working, and runtime dependencies - remains valuable for understanding the 13.5× average expansion from declared to runtime dependencies in successful projects, but the primary challenge is generating syntactically and logically correct code.

Key findings include dramatic language differences (Python 89.2\% vs Java 44.0\% success), unexpected agent specializations (Gemini's perfect Python, Claude's Java expertise), and universal struggles with dependency transitivity. These results demonstrate that while LLMs generate syntactically correct code, they fail to capture the full execution context required for reproducibility.

The implications are significant for multiple stakeholders. Developers spend substantial time debugging AI-generated code approximately 15 minutes per failed project. Organizations should select agents based on technology stacks rather than assuming uniform performance. The research community needs benchmarks that evaluate executable reliability, not just functional correctness. Most critically, the 31.7\% of projects requiring manual intervention represents a hidden cost that must be factored into productivity assessments of AI coding agents.

Our study exposes an inconvenient truth: AI coding agents generate code that looks complete but isn't reproducible. The 31.7\% failure rate and 13.5× dependency gap aren't just statistics they represent thousands of hours developers waste debugging code that should have worked. Until LLMs learn to specify complete execution environments, not just code logic, they remain sophisticated autocomplete tools rather than true development partners.

The path forward is clear but challenging. LLMs must train on complete projects with full dependency chains, not code fragments. They need to understand that importing scikit-learn means 52 packages at runtime, not just one line in requirements.txt. Most importantly, they should test their own generated code in clean environments, catching reproducibility failures before developers encounter them.

We've quantified the reproducibility crisis in AI-generated code and provided a framework for measuring it. The question now isn't whether these tools can write code they clearly can. The question is whether the software development community will demand reproducible code from AI systems, or accept the hidden tax of manual debugging as the price of automation. Only when reproducibility becomes a primary metric, not an afterthought, will AI truly accelerate software development.

\bibliography{aaai2026}

@article{gundersen2018on,
  author  = "Gundersen, Odd Erik and Gil, Yolanda and Aha, David W.",
  year    = 2018,
  title   = "{On Reproducible AI: Towards Reproducible Research, Open Science, and Digital Scholarship in AI Publications}",
  journal = "AI Magazine",
  volume  = 39,
  number  = 3,
  pages   = "56--68",
  doi     = "10.1609/aimag.v39i3.2816",
  publisher = "Association for the Advancement of Artificial Intelligence"
}

@article{pineau2021improving,
  author  = {Pineau, Joelle and Vincent-Lamarre, Philippe and Sinha, Koustuv and Larivi{\`e}re, Vincent and Beygelzimer, Alina and d’Alch{\'e}-Buc, Florence and Fox, Emily B. and Larochelle, Hugo},
  year    = {2021},
  title   = {Improving Reproducibility in Machine Learning Research (A Report from the NeurIPS 2019 Reproducibility Program)},
  journal = {Journal of Machine Learning Research},
  volume  = {22},
  number  = {164},
  pages   = {1--20},
  url     = {https://jmlr.org/papers/v22/20-303/20-303.pdf},
}

@misc{jain2024livecodebench,
  author       = {Jain, Naman and Han, King and Gu, Alex and Li, Wen-Ding and Yan, Fanjia and Zhang, Tianjun and Wang, Sida and Solar-Lezama, Armando and Sen, Koushik and Stoica, Ion},
  title        = {LiveCodeBench: Holistic and Contamination Free Evaluation of Large Language Models for Code},
  year         = {2024},
  howpublished = {arXiv preprint arXiv:2403.07974},
  url          = {https://livecodebench.github.io/}
}

@smisc{reprozip,
  author       = {Chirigati, Fernando and Rampin, Rémi and others},
  year         = {2024},
  title        = {ReproZip: A tool to bundle and reproduce computational experiments},
  publisher    = {VIDA Lab, New York University},
  url          = {http://docs.reprozip.org/},
  note         = {Accessed: 2025-10-20}
}

@INPROCEEDINGS{8109156,
  author={Ton That, Dai Hai and Fils, Gabriel and Yuan, Zhihao and Malik, Tanu},
  booktitle={2017 IEEE 13th International Conference on e-Science (e-Science)}, 
  title={Sciunits: Reusable Research Objects}, 
  year={2017},
  volume={},
  number={},
  pages={374-383},
  doi={10.1109/eScience.2017.51}
}

@article{wang2025assessing,
  title={Assessing consistency and reproducibility in the outputs of large language models: Evidence across diverse finance and accounting tasks},
  author={Wang, Julian Junyan and Wang, Victor Xiaoqi},
  journal={arXiv preprint arXiv:2503.16974},
  year={2025}
}

@inproceedings{staudinger2024reproducibility,
  title={A reproducibility and generalizability study of large language models for query generation},
  author={Staudinger, Moritz and Kusa, Wojciech and Piroi, Florina and Lipani, Aldo and Hanbury, Allan},
  booktitle={Proceedings of the 2024 Annual International ACM SIGIR Conference on Research and Development in Information Retrieval in the Asia Pacific Region},
  pages={186--196},
  year={2024}
}

@article{chen2021evaluating,
  author    = {Chen, Mark and Tworek, Jerry and Jun, Heewoo and Yuan, Qiming and Ponde de Oliveira Pinto, Henrique and Kaplan, Jared and Edwards, Harri and Burda, Yuri and Joseph, Nicholas and Brockman, Greg and Ray, Alex and Puri, Raul and Krueger, Gretchen and Petrov, Michael and Khlaaf, Heidy and Sastry, Girish and Mishkin, Pamela and Chan, Brooke and Gray, Scott and Ryder, Nick and ... and Sutskever, Ilya and Zaremba, Wojciech},
  title     = {Evaluating Large Language Models Trained on Code},
  journal   = {arXiv preprint arXiv:2107.03374},
  year      = {2021},
  url       = {https://arxiv.org/abs/2107.03374}
}

@inproceedings{yang2024swe,
  title={SWE-bench: Can Language Models Resolve Real-world GitHub Issues?},
  author={Yang, John and Jimenez, Carlos E and Leblond, Alexander and Hoppe, Bernardo and Tian, Katherine and Taori, Rohan and Hu, Edward J and Zhang, Hao and Koh, Pang Wei and Hudson, Drew A and others},
  booktitle={International Conference on Learning Representations},
  year={2024},
  note={Notes that environment configuration remains a bottleneck for LLM-based software engineering agents}
}

@misc{anthropic2025claude,
  title={Claude Opus 4.1: Advanced Code Generation Model},
  author={{Anthropic}},
  year={2025},
  month={August},
  day={5},
  howpublished={Model release announcement},
  url={https://www.anthropic.com/claude}
}

@misc{openai2025codex,
  title={OpenAI Codex CLI Version 0.52.0},
  author={{OpenAI}},
  year={2025},
  howpublished={Command-line tool release},
  url={https://platform.openai.com/docs/guides/code}
}

@misc{google2025gemini,
  title={Gemini 2.5 Pro: Enhanced Code Generation Capabilities},
  author={{Google DeepMind}},
  year={2025},
  month={June},
  day={17},
  howpublished={Model release announcement},
  url={https://deepmind.google/technologies/gemini}
}

@misc{stackoverflow2025,
  title={Stack Overflow Developer Survey 2025},
  author={{Stack Overflow}},
  year={2025},
  howpublished={Annual developer survey},
  note={Reports JavaScript usage at 66\%, Python at 57.9\%, and Java at 29.4\% among all respondents},
  url={https://survey.stackoverflow.co/2025}
}

@misc{jetbrains2024,
  title={The State of Developer Ecosystem 2024},
  author={{JetBrains}},
  year={2024},
  howpublished={Annual developer ecosystem report},
  url={https://www.jetbrains.com/lp/devecosystem-2024}
}

@inproceedings{sculley2015hidden,
  title={Hidden Technical Debt in Machine Learning Systems},
  author={Sculley, David and Holt, Gary and Golovin, Daniel and Davydov, Eugene and Phillips, Todd and Ebner, Dietmar and Chaudhary, Vinay and Young, Michael and Crespo, Jean-Fran{\c{c}}ois and Dennison, Dan},
  booktitle={Advances in Neural Information Processing Systems},
  volume={28},
  year={2015},
  note={Seminal paper on technical debt in ML systems}
}

@article{sweller1988cognitive,
  title={Cognitive Load During Problem Solving: Effects on Learning},
  author={Sweller, John},
  journal={Cognitive Science},
  volume={12},
  number={2},
  pages={257--285},
  year={1988},
  publisher={Elsevier},
  note={Foundational paper on Cognitive Load Theory}
}

@article{raff2019step,
  title={A step toward quantifying independently reproducible machine learning research},
  author={Raff, Edward},
  journal={Advances in Neural Information Processing Systems},
  volume={32},
  year={2019}
}

@article{hutson2018artificial,
  title={Artificial intelligence faces reproducibility crisis},
  author={Hutson, Matthew},
  journal={Science},
  volume={359},
  number={6377},
  pages={725--726},
  year={2018}
}

@inproceedings{haibe2020transparency,
  title={Transparency and reproducibility in artificial intelligence},
  author={Haibe-Kains, Benjamin and others},
  booktitle={Nature},
  volume={586},
  pages={E14--E16},
  year={2020}
}

@inproceedings{austin2021program,
  title={Program synthesis with large language models},
  author={Austin, Jacob and others},
  booktitle={arXiv preprint arXiv:2108.07732},
  year={2021}
}

@article{jimenez2024swebench,
  title={SWE-bench: Can Language Models Resolve Real-World GitHub Issues?},
  author={Jimenez, Carlos E and others},
  journal={arXiv preprint arXiv:2310.06770},
  year={2024}
}

@inproceedings{decan2019empirical,
  title={An empirical comparison of dependency network evolution in seven software packaging ecosystems},
  author={Decan, Alexandre and Mens, Tom and Grosjean, Philippe},
  booktitle={Empirical Software Engineering},
  volume={24},
  pages={381--416},
  year={2019}
}

@article{kikas2017structure,
  title={Structure and evolution of package dependency networks},
  author={Kikas, Riivo and others},
  journal={IEEE/ACM Working Conference on Mining Software Repositories},
  pages={102--112},
  year={2017}
}

@article{peng2011reproducible,
  title={Reproducible research in computational science},
  author={Peng, Roger D},
  journal={Science},
  volume={334},
  number={6060},
  pages={1226--1227},
  year={2011}
}


\end{document}